\documentclass[12pt,authoryear]{article}

\title{US-ELTP-WhitePaper}
\usepackage{lscape}
\usepackage[english]{babel}
\usepackage{fancyhdr}
\usepackage{eurosym}
\usepackage{tcolorbox}

\usepackage{xcolor}
\definecolor{bluegray}{rgb}{0.4, 0.6, 0.9}
\usepackage{multirow}
\usepackage{multicol}
\usepackage{array}
\newcolumntype{L}[1]{>{\raggedright\arraybackslash}p{#1}}
\usepackage{titlesec}
\usepackage{enumerate}
\usepackage{longtable}
\usepackage{mathptmx}
\usepackage{graphicx}
\usepackage{epsfig}
\usepackage{sidecap}
\usepackage{amssymb}
\usepackage{hyperref} 
\usepackage[final]{pdfpages}
\usepackage{wrapfig}
\usepackage{wasysym}
\usepackage{setspace} 
\usepackage{natbib}
\usepackage{enumitem}
\RequirePackage{lscape}
\renewcommand{\bibpreamble}{\begin{multicols}{2}}
\renewcommand{\bibpostamble}{\end{multicols}}
\newlength{\bibitemsep}
\newlength{\bibparskip}
\let\oldthebibliography\thebibliography
\renewcommand\thebibliography[1]{%
  \oldthebibliography{#1}%
  \setlength{\parskip}{\bibitemsep}%
  \setlength{\itemsep}{\bibparskip}%
}

\def\hawaii{Hawai$\!$`i}

\begin {document}

\pagenumbering{roman}


\vspace{1.5cm}
\noindent
\\
\noindent{\Large \bf \textcolor{black}\noindent{Transformative Planetary Science with the US ELT Program}}

\begin{figure}[h!]
\centerline{\includegraphics[width=17.0cm]{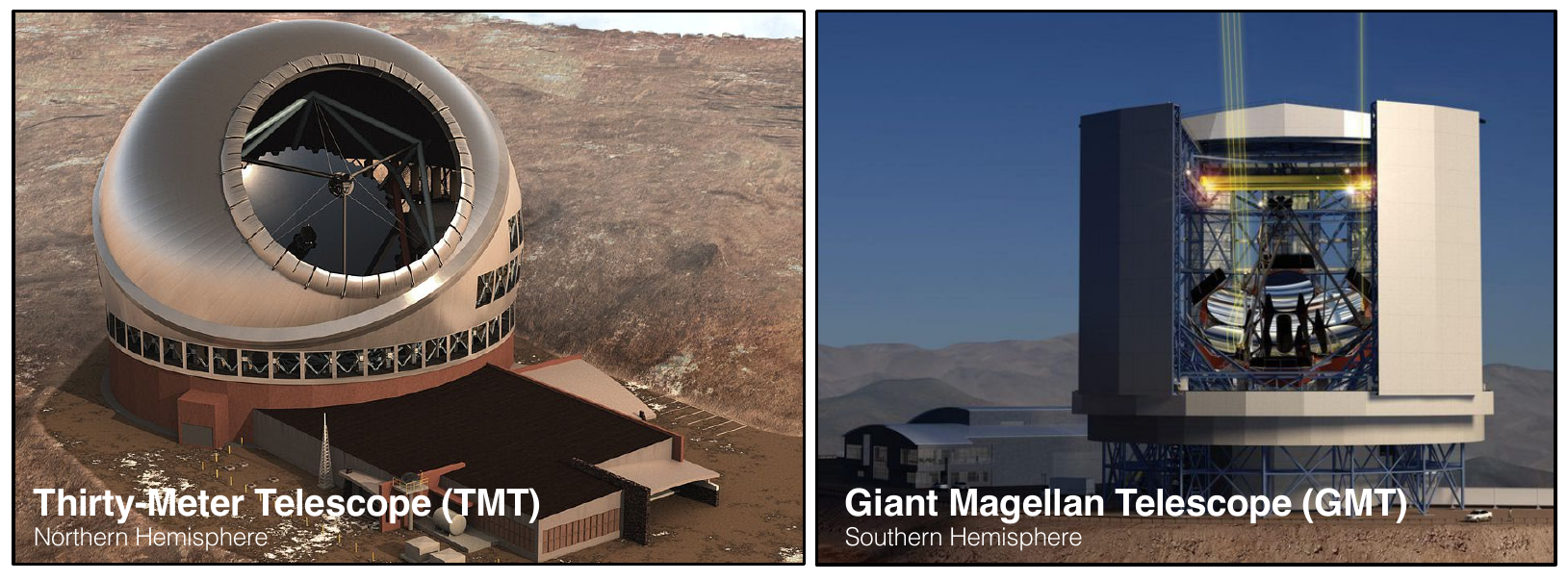}}
\end{figure}

\vspace{0.2cm}
\noindent
{\bf Primary Author:} \\
Michael H. Wong (UC Berkeley / SETI Institute) \\ +1-510-224-3411,   \textit{mikewong@astro.berkeley.edu} \\

\vspace{0.1cm}
\noindent
{\bf Lead Authors:} \\
Karen J. Meech (University of \hawaii, Institute for Astronomy), \textit{meech@hawaii.edu} \\
Mark Dickinson (NSF's NOIRLab), \textit{med@noao.edu}\\
Thomas Greathouse (Southwest Research Institute), \textit{tgreathouse@swri.edu}\\
Richard J. Cartwright (SETI Institute), 
\textit{rcartwright@seti.org} \\
Nancy Chanover (New Mexico State University), \textit{nchanove@nmsu.edu} \\
Matthew S. Tiscareno (SETI Institute),
\textit{matt@seti.org}\\

\vspace{0.1cm}
\noindent
{\bf Endorsers:}\\
 Tracy Becker (SwRI),
 Susan Benecchi (PSI), 
 Gordon L. Bjoraker (NASA GSFC), 
 Ben Byron (SwRI), 
 Al Conrad (LBTO), 
 Imke de Pater (UC Berkeley), 
 Joshua P. Emery (NAU), 
 Jonathan Fortney (UC Santa Cruz), 
 Patrick Fry (Univ. Wisconsin), 
 Rohini Giles (SwRI), 
 Randy Gladstone (SwRI), 
 Caitlin Griffith (Univ.\ of Arizona LPL), 
 Bryan Holler (Space Telescope Science Institute), 
 Vincent Hue (SwRI), 
 Hope Ishii (Univ. \hawaii), 
 Joshua Kammer (SwRI), 
 Jan Kleyna (Univ. \hawaii), 
 Michael P. Lucas (Univ. of Tennessee), 
 Melissa McGrath (SETI Institute), 
 Erica Molnar-Bufanda (Univ. \hawaii), 
 Philippa Molyneux (SwRI), 
 Marc Neveu (Univ. of Maryland, NASA GSFC), 
 Conor Nixon (NASA GSFC),
 Tom A. Nordheim (JPL), 
 Glenn S. Orton (JPL), 
 Angel Otarola (ESO), 
 N. Pinilla-Alonso (FSI/UCF), 
 Shalima Puthiyaveettil (Manipal Academy of Higher Education), 
 Ujjwal Raut (SwRI), 
 Kurt Retherford (SwRI), 
 Matthew Richter (UC Davis), 
 Megan E. Schwamb (Queen's Univ. Belfast), 
 Driss Takir (JETS/ARES, NASA Johnson), 
 David Trilling (Northern Arizona University), 
 Christian Veillet (LBTO), 
 Geronimo Villanueva (NASA GSFC), 
 Richard Wainscoat (Univ. \hawaii), 
 Xi Zhang (UC Santa Cruz)
\newpage

\pagenumbering{arabic}

\vspace{-1cm}
\begin{tcolorbox}[colback=blue!5!white,colframe=black!75!black]
\vspace{-0.2cm}
\section{Executive Summary}
\label{sec:intro}
\vspace{-0.2cm}


\textit{The proposed US Extremely Large Telescope (ELT) Program would secure national open access to at least 25\% of the observing time on the Thirty Meter Telescope in the north and the Giant Magellan Telescope in the south. ELTs would advance solar system science via exceptional angular resolution, sensitivity, and advanced instrumentation. ELT contributions would include the study of interstellar objects, giant planet systems and ocean worlds, the formation of the solar system traced through small objects in the asteroid and Kuiper belts, and the active support of planetary missions. We recommend that (1) the US ELT Program be listed as critical infrastructure for solar system science, that (2) some support from NASA be provided to ensure mission support capabilities, and that (3) the US ELT Program expand solar-system community participation in development, planning, and operations.
}
\end{tcolorbox}


\vspace{-0.5cm}
\section{The US ELT Program} 
\label{sec:Overview}

\vspace{-0.3cm}
\subsection{Program Overview} 
\label{sec:ELT-overview}
\vspace{-0.2cm}

Over the last 25 years, a generation of 8m-class optical-infrared telescopes has allowed us to study faint objects from the outer reaches of our solar system to the first billion years of cosmic history. Over the same period, the promise of adaptive optics (AO) to correct the blurring effects of Earth’s atmosphere has been realized.  Today, some of the most important problems in astrophysics and planetary science require observations with higher angular resolution and sensitivity than current telescopes can provide \citep{skidmore2015,bernstein2018}. A new generation of extremely large telescopes (ELTs) is technically achievable and scientifically compelling. For diffraction-limited observations, sensitivity\footnote{Here, sensitivity is defined as the reciprocal of the exposure time required to attain a given signal-to-noise ratio for observations of a point source with a given flux in the background-limited regime.} scales with primary mirror diameter ($D$) as $D^4$, or even more steeply for observations in crowded fields, so the potential gains from telescopes with $D > 20$~m are enormous.

Three international ELT projects are underway, including two with substantial leadership from US research institutions: the 24.5m Giant Magellan Telescope (GMT) in the southern hemisphere and the Thirty Meter Telescope (TMT) in the northern hemisphere.  The European Southern Observatory (ESO) is building a 39m Extremely Large Telescope in Chile.

The US ELT Program (US-ELTP)\footnote{\url{https://www.noirlab.edu/public/projects/useltp/}; \url{https://www.noao.edu/us-elt-program/}}  is a joint endeavor of NSF’s NOIRLab and the organizations building the TMT and GMT \citep{wolff2019}. The program goal is to complete both telescopes and to secure sufficient federal funding to make at least 25 percent of the observing time available for open access by the whole US scientific community. This two-telescope, two-hemisphere ELT system would provide the US community with greater and more diverse research opportunities than could be achieved with a single telescope, while creating new channels for collaboration with scientists  within GMT’s and TMT’s international partners (Japan, India, China, Canada, Australia, Brazil, and Korea). The US-ELTP would  enable large-scale, systematic investigations of forefront scientific problems (Key Science Programs, Sec.~\ref{sec:ELT-SS-community}) that could not be achieved with smaller, uncoordinated projects. NOIRLab will develop and provide user support services and tools for all stages of research with TMT and GMT, from preparing observing proposals to analyzing the data, as well as a permanent archive for all GMT/TMT data. The proposed federal investment would maximize the scientific return from a cooperative, national US ELT Program.

\vspace{-0.3cm}
\subsection{Unique Capabilities for Solar System Science} 
\label{sec:ELT-capabilities}
\vspace{-0.2cm}



\begin{figure}[ht!]
\centerline{\includegraphics[width=15.5cm]{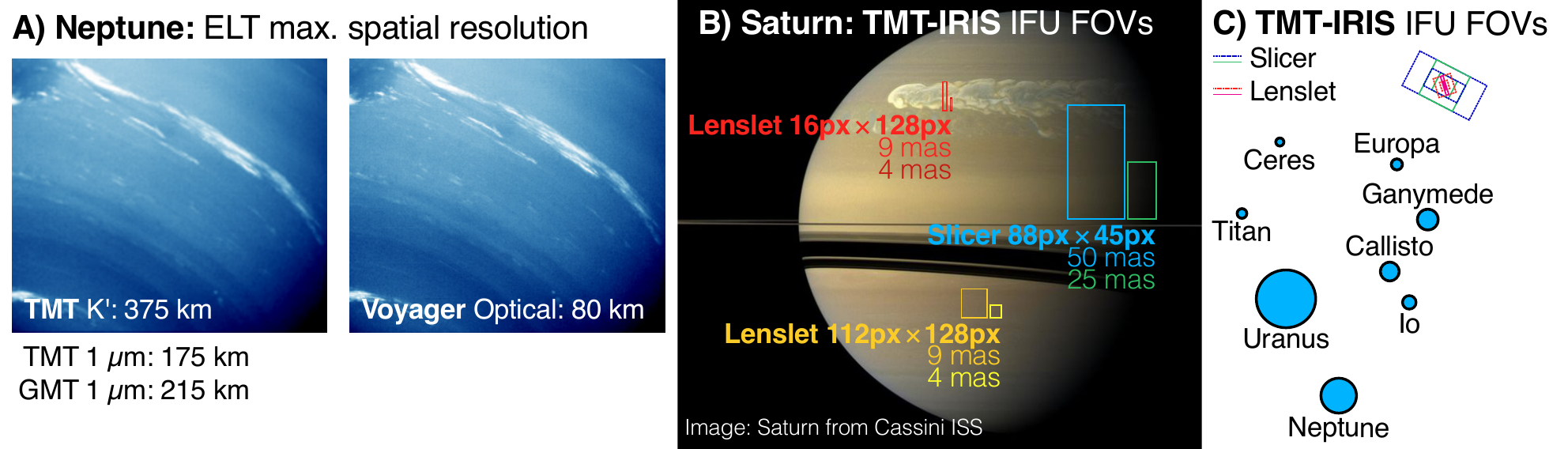}}
\vspace{-0.4cm}
\caption{\it ELTs will be capable of spacecraft-mission level science. \textbf{A)} At 1 $\mu$m, the diffraction-limited TMT (GMT) spatial resolution will be 11 (13)
km at Ceres and 175 (215) km at Neptune. \textbf{B)} Hyperspectral datasets can resolve inhomogeneous regions that would be blurred together by current instrumentation. \textbf{C)} Global spectral mapping can be efficiently achieved on numerous outer solar system bodies with the TMT-IRIS first-generation spectro-imager.}
\label{fig:FOVs}
\end{figure}

The angular resolution of ELTs operating with adaptive optics, as well as their sensitivity, will enable transformational new research on objects throughout the solar system.  ELTs will provide optical-infrared imaging and spectroscopy with angular resolution previously achievable only by fly-by and rendezvous missions (Fig.~\ref{fig:FOVs}).  They offer the opportunity for long-duration observing campaigns, and for fast response to observe time-critical events. The two-telescope, two-hemisphere US ELT Program would offer all-sky coverage for the US scientific community, particularly important for high-inclination targets like interstellar objects (ISOs) (Sec.~\ref{sec:ISO}) and long-period comet populations. The longitudinal separation of GMT and TMT will enable continuous observations of targets near the ecliptic over a longer duration than is possible at one site, particularly important for targets with short rotational or orbital periods.

\vspace{-0.15cm}

\vspace{-0.25cm}

In addition to solar system science, the US-ELTP has enormous relevance for studies of exoplanetary systems and exoplanet atmospheres, including comparative planetology and the search for biosignatures.  These topics are addressed in many white papers submitted to Astro2020.\footnote{\url{https://www.noao.edu/us-elt-program/astro2020.php}}

\vspace{-0.3cm}
\subsection{Solar System Community Involvement} 
\label{sec:ELT-SS-community}
\vspace{-0.2cm}

The research community has given extensive consideration to the scientific potential of ELTs for studying objects in our solar system.  For example, the TMT Detailed Science Case \citep{skidmore2015} contains a chapter on solar system science, and community scientists submitted several science white papers on solar system research using ELTs to the Astro2020 Decadal Survey \citep{bauer2019,chanover2019,meech2019,trilling2019,wong2019}.  ELT-related workshops have been held at solar system science conferences (in addition to the typical astronomy conferences). For example, over 50 participants registered for a TMT workshop at the AAS-DPS meeting in 2015. Solar system scientists also have roles as TMT observatory technical and management staff, in addition to advisory roles for science definition and instrument development. In a 2018 NOIRLab community planning exercise,  researchers developed concepts for several US-ELTP Key Science Programs in solar system astronomy. A national US ELT Program will enable solar system scientists at all US research institutions to use these next-generation observatories and to contribute to planning their futures.





\vspace{-0.5cm}
\section{Planetary Science Cases}
\label{sec:planetary}


\vspace{-0.3cm}
\subsection{Interstellar Objects} 
\label{sec:ISO}
\vspace{-0.2cm}

Small primitive bodies can provide information about the solar system’s formative processes, including the contribution of presolar and interstellar sources. ISOs are thought to be planetesimal remnants that have been ejected out of their own solar system \citep{raymond2018}, and they can provide insight into the process of building exoplanetary systems.

The recent discovery of the first ISO, 1I/’Oumuamua \citep{meech2017} provoked intense, sustained interest in the scientific community although there was only a very brief period during which it was possible to observe this ISO as it moved rapidly away from the Earth.  ’Oumuamua did not look like a normal comet. While some characteristics were typical of comets, others were not. 
Much of the science that the community wanted to address, notably chemistry, was impossible without a much larger facility such as the US-ELTP. The second ISO discovered in 2019, C/2019 Q4 (Borisov), was quite different, behaving like a comet driven primarily by CO-sublimation \citep{bodewits2020,cordiner2020}. 

These two objects illustrate the opportunity to assess similarities and differences in the chemistry and physical processes driving planetary growth in other planetary systems.  The study of interstellar objects incorporates a broad range of scientific disciplines including galactic, stellar, and planetary dynamics, planetesimal formation, tidal disruption, shape modeling and the nature and evolution of cometary nuclei.  Finding a large number of ISOs will be made possible by the Rubin Telescope, where discovery rates are predicted to exceed one per year \citep{trilling2017b}. If ‘Oumuamua’s size is representative of most ejected planetesimals, fully characterizing these will require the use of space telescopes and the next generation of extremely large telescopes.  To date, there have been over 100 papers written on these first two ISOs ranging from characterization, planetesimal formation, stellar and galactic dynamics, tidal disruption and shape etc. The discovery an follow up of ISOs has stimulated a remarkable burst of science, and in the next decade we need to be ready to take advantage of what can be learned from these objects as a class.

\vspace{-0.3cm}
\subsection{Testing Solar System Formation Models -- Small Solar System Bodies} 
\label{sec:Manx}
\vspace{-0.2cm}

The most primitive Solar System objects are found beyond the asteroid belt. They have largely been undisturbed since since formation. Thus, their compositions provide a fossilized record of the chemical make-up of our planetary system during its origin. Larger bodies formed from the accumulation of many smaller bodies. This accretionary process had geophysical and geochemical consequences as material heated and differentiated, thereby erasing the signature of their primordial compositions. The most primitive of these early planetesimals are likely to be the smallest which have not been thermally processed, thus they preserve a chemical fingerprint of the disk chemistry. Because these are the faintest, they are also the most difficult to physically characterize.

Kuiper belt objects are rocky/icy bodies that orbit outside the orbit of Neptune. Centaurs, with semi-major axes between the orbits of Jupiter and Uranus, are thought to be relatively recent ($<$10 Myr) ``escapees'' from the Kuiper belt on their way to becoming short period comets. They provide convenient compositional markers between comets and their original reservoir. Comets are km-scale volatile-rich bodies that formed outside the solar system's snowline. There are several dynamical reservoirs today for comets which can trace conditions at different locations in the disk. 

\vspace{-0.3cm}
\subsubsection{Main Belt Comets}
\vspace{-0.2cm}
No one knows how water arrived at our planet, or whether our solar system, with a planet possessing the necessary ingredients for life within the habitable zone, is a cosmic rarity. We do not know the role that the gas giants played in delivering essential materials to the habitable zone. The answers to these questions are contained in unaltered primitive body volatiles. The most likely volatile delivery candidate came from the outer solar system \citep{meech2020}.  

Evidence of what arrived at Earth is hard to determine from Earth materials because of the processing that has occurred.  However, whatever was delivered to the inner solar system was also trapped in the asteroid belt where some of those primitive materials exist today. Studying accessible volatiles in the main belt, found in the main belt comets (MBCs) will provide the key information needed to understand the origins of these materials.  However the detailed isotopic work needed requires in-situ exploration, and the ELTs will provide the necessary groundwork. Currently, direct spectroscopic detection of volatiles is not possible with Earth-based telescopes \citep{snodgrass2017} for MBCs because of their small sizes (radii $<$ 2 km) and low expected outgassing rates.  Thirty-meter class facilities (or potentially the James Webb Space Telescope) may be able to directly characterize the chemical composition of MBC volatiles.

\begin{figure}[h!]
\vspace{-0.1cm}
\centerline{\includegraphics[width=15.0cm]{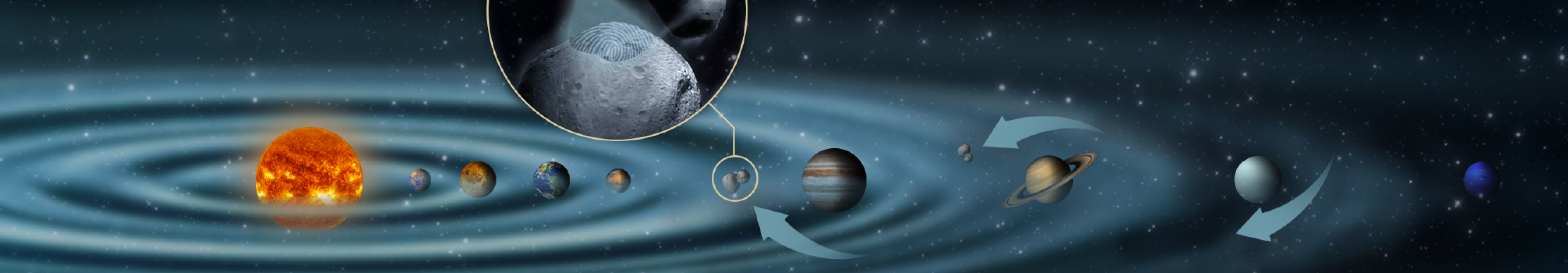}}
\vspace{-0.4cm}
\end{figure}

\vspace{-0.3cm}
\subsubsection{Manx Comets}
\vspace{-0.2cm}

Manx comets are a new class of objects coming in from the Oort cloud which exhibit little or no comet activity \citep{meech2016}. Studying Manx surface mineralogy will trace how things were dynamically scattered during the formation process, testing solar system dynamical models.  Different models make different predictions on the relative distances and thus compositions of planetesimals that were scattered out to the Oort cloud that we can sample now as Manxes. Distinguishing between small body surface composition classes is straightforward. Diagnostic ice and organic features for minerals require only low-resolution near-IR spectra (R$<$1,000; 0.8-2.4 $\mu$m). This is difficult with 8-10 m telescopes for faint objects with short visibility periods, but ELT sensitivity could reach the signal to noise ratios needed to distinguish solar system models.

\vspace{-0.3cm}
\subsubsection{Centaurs and Kuiper Belt Objects}
\vspace{-0.2cm}

ELTs will produce revolutionary insight into the structure, properties, and evolution of minor bodies by providing order(s) of magnitude improvements with disk-resolved maps. Disk-resolved imagery of outer Solar System objects offers a completely new view on their evolution which can probe the formation mechanisms in the outer Solar System. Instruments with diffraction limited imaging and spectroscopic capabilities can study the shapes and compositions of Centaurs, resolving complex structures such as the rings around Chariklo \citep{braga2014}. Large ground-based telescopes can produce near-spacecraft quality geologic maps of the largest asteroids \citep{fetick2019}. Candidate ocean worlds Pluto and Triton \citep{hendrix2019}, along with Charon and Arrokoth, are the only KBOs whose surfaces have been spatially resolved. Spectral maps from ELTs
could be used to study surface properties of these large primordial bodies, and measure
changes from any ongoing interior and atmospheric processes.

\vspace{-0.3cm}
\subsection{Giant Planet Systems} 
\label{sec:gp-toplevel}

\vspace{-0.3cm}
\subsubsection{Satellites and Ocean Worlds}
\vspace{-0.2cm}
Some of the most likely places where life might exist beyond Earth are in subsurface liquid H$_2$O oceans contained within the interiors of icy moons. Confirmed `ocean worlds' include Europa, Ganymede, Callisto, Enceladus and Titan, and other, possible ocean worlds include Neptune's largest moon Triton, the five mid-sized moons of Uranus, and Saturn's other mid-sized moons \citep{hendrix2019}. Evidence for ongoing plume activity has been detected on Europa \citep{roth2014} and Enceladus \citep{porco2006}, strong evidence for recent plume activity has been detected on Triton \citep{smith1989}, and hints of cryovolcanism in the recent past have been detected on Uranus' moons Miranda and Ariel \citep{smith1986}. 

The surfaces of these confirmed and possible ocean worlds are rich in ices, organics, salts, and other geologically short-lived minerals. Many of these constituents are thermodynamically unstable over seasonal timescales, and they sublimate into tenuous atmospheres where they are subsequently transported to cold traps and condense. As a result, icy moon surface compositions are constantly changing as volatile species are transported across their surfaces. Long-term observing programs are therefore crucial for tracking how the surfaces of icy moons change over time \citep{cartwright2020,holler2016}. The superior angular resolution and sensitivity that can be achieved by ELTs would dramatically expand on previous efforts, allowing us to better identify and characterize trends in the changing spatial distribution of volatiles on icy moons. 

Furthermore, ELTs could monitor ongoing volcanic activity on Io, which supplies material to the plasma torus, subsequently modifying and enriching the surface chemistries of the icy Galilean moons Europa, Ganymede, and Callisto. Additionally, long-term monitoring of the dwarf planet Ceres, another possible ocean world that exhibits evidence for recent endogenic activity, could be conducted by ELTs to monitor changes in its surface composition \citep{castillo-rogez++2020}.

\vspace{-0.3cm}
\subsubsection{Atmospheric Dynamics}
\label{sec:gp-dynamics}
\vspace{-0.2cm}

Giant planet atmospheric observations require a range of observing cadences. Some dynamic features---such as asteroid/comet impacts and superstorm eruptions---are generated unpredictably and require rapid response observations. Impacts provide science insights ranging from the nature and size distribution of impactor populations, to stratospheric chemical and dynamical processes that evolve over timescales from minutes to even centuries \citep{harrington++2004,hammel++2010,hueso++2013,moreno++2017}. Giant storms trace heat transport within hydrogen-dominated atmospheres, particularly the nature of moist convection and its inhibition by molecular weight stratification. We are learning now that the radiative-convective balance in Uranus and Neptune may be governed by this very non-earthlike convective regime, such that the zero thermal flux of Uranus derived by Voyager may have represented a quiescent inter-storm period, while much of the heat flux may be carried by episodic storms \citep{smith+gierasch1995,friedson+gonzales2017}.

ELTs will lead to quantitative advances because they will enable spectroscopy of compact regions with wildly variable properties (Fig.~\ref{fig:FOVs}), from impact debris fields to active storms. The data will spatially resolve variation in gas composition and aerosol properties, which in turn trace chemical and physical processes related to convection and thermal evolution that are ultimately related to our interpretation of giant exoplanet observations \citep{fortney+nettelmann2010}.

Mid-infrared capability at diffraction-limited resolution, although not planned for first
light with TMT or GMT, will advance our understanding of ice giant climate and atmospheric dynamics to the level we
have for Jupiter and Saturn. 

\vspace{-0.3cm}
\subsubsection{Auroral Emission}
\label{sec:gp-aurora}
\vspace{-0.2cm}
In giant planet ionospheres, emission from H$_3^+$ in the 2--4 $\mu$m region reveals interactions with the deeper atmosphere (via chemical transport and wave breaking) and with the magnetosphere (via auroral precipitaion). Ground-based observations have demonstrated rapid variation in precipitating electron fluxes at Jupiter \citep{watanabe++2018}, traced solar wind interactions at Jupiter and Saturn \citep{stallard++2012,sinclair++2017}, and detected airglow and possible auroral emission at Uranus \citep{melin++2019}. At Neptune, the higher sensitivity of ELTs may enable a first detection of H$_3^+$ emission, which is much lower than predicted by models \citep{melin++2018}. At Jupiter, higher spatial resolution would allow ground-based observations to reach higher latitudes, and would enabled detailed studies of auroral features such as satellite footprints and correspondence between auroral emission and auroral-related heating of the neutral atmosphere.  

\vspace{-0.3cm}
\subsubsection{Rings}
\label{sec:gp-rings}
\vspace{-0.2cm}
Planetary rings serve as accessible natural laboratories for disk processes, as clues to the origin and evolution of planetary systems, and as shapers and detectors of their planetary environments \citep{tiscareno+murray2018}. High spatial resolution is needed to characterize the evolution of the ring arcs of Neptune, apparently Myr-unstable inner moons of Uranus, spokes and propellers in Saturn's rings, arcs in the rings of Jupiter, and moons with poorly understood orbital variations such as Daphnis and Prometheus. Evolving wave structures within planetary rings, potentially resolvable with ELTs and future space-based optical telescopes, record the impact history in the Jupiter \citep{showalter++2011} and Saturn \citep{hedman++2015} systems, as well as internal oscillations that constrain the deep structure of gas giants \citep{mankovich++2019}.

\vspace{-0.3cm}
\subsection{Spacecraft Mission Support}
\label{sec:missionsupport}
\vspace{-0.2cm}

NASA’s exploration of our Solar System relies upon both well-instrumented spacecraft sent to sample and make up close observations of targets such as planets, moons, asteroids and comets, and a rich and diverse array of ground-based and Earth orbiting observatories.  
Space missions achieve viewing geometries often impossible from Earth, while Earth observatories have little to no limitations on data volume or instrument/observatory size.  Synergy between these types of facilities has expanded our understanding of the formation and evolution of our solar system in ways no one observatory or mission could ever do on their own.  The need for ground based support of NASA missions is demonstrated by NASA’s continued use of the NASA Infrared Telescope Facility (IRTF) and the NASA/Keck partnership.  The NASA/Keck time is dominated by mission support, much of which has been spent supporting  Cassini, New Horizon, Juno, and the future Europa Clipper mission to name a few. Mission support activities include hazard avoidance and mission target discovery \citep{showalter+hamilton2015,porter++2018,buie++2020}.

NASA missions under development such as Europa Clipper and Dragonfly will be performing detailed studies of icy moons at Jupiter and Saturn along with ESA's JUICE mission, with its NASA instrumentation, to study Ganymede.  The ELTs' incredible AO corrected spatial resolution could be leveraged to increase temporal imaging and high-resolution spectroscopic observations of these moons and to track the activity occurring on neighboring moons, responsible for loading the magnetospheres with plasma.  The ELTs could spatially resolve individual plumes on Io, as a case in point  \citep{skidmore2015}.  Not only will the ELTs support the science performed with these and future NASA missions, but the unique imagery they produce will serve to keep the public engaged and excited in NASA’s exploration of our solar system for years to come.  

\vspace{-0.4cm}
\bibliographystyle{aasjournal}
\bibliography{references}{}

\end{document}